\documentclass[sigconf,natbib=true,authordraft,review=false]{acmart}
\AtBeginDocument{%
  \providecommand\BibTeX{{%
    \normalfont B\kern-0.5em{\scshape i\kern-0.25em b}\kern-0.8em\TeX}}}

\usepackage{caption,nccmath}
\usepackage{subcaption}
\usepackage[titlenumbered,ruled]{algorithm2e}
\usepackage{multirow}
\usepackage{arydshln}
\usepackage{amsmath}

\usepackage{cleveref}

\begin{document}




\title{Causal Disentangled Variational Auto-Encoder for Preference Understanding in Recommendation}

\author{Siyu Wang}
\affiliation{%
  \institution{The University of New South Wales}
  \city{Sydney}
  \state{NSW}
  \country{Australia}
  \postcode{2052}
}
\orcid{0009-0008-8726-5277}
\email{siyu.wang5@student.unsw.edu.au}

\author{Xiaocong Chen}
\affiliation{%
  \institution{The University of New South Wales}
  \city{Sydney}
  \country{Australia}
}
\email{xiaocong.chen@unsw.edu.au}

\author{Quan Z. Sheng}
\affiliation{%
 \institution{Macquarie University}
 \city{Sydney}
  \country{Australia}
  }
\email{michael.sheng@mq.edu.au}

\author{Yihong Zhang}
\affiliation{%
 \institution{Osaka University}
 \city{Osaka}
  \country{Japan}
  }
\email{zhang.yihong@ist.osaka-u.au.jp}

\author{Lina Yao}
\affiliation{%
  \institution{Data61, CSIRO}
  \city{Eveleigh}
  \country{Australia}
}
\affiliation{%
  \institution{The University of New South Wales}
  \city{Sydney}
  \country{Australia}
}
\email{lina.yao@unsw.edu.au}

\renewcommand{\shortauthors}{Wang, et al.}

\begin{abstract}
Recommendation models are typically trained on observational user interaction data, but the interactions between latent factors in users' decision-making processes lead to complex and entangled data. Disentangling these latent factors to uncover their underlying representation can improve the robustness, interpretability, and controllability of recommendation models.
This paper introduces the \textbf{Ca}usal \textbf{D}isentangled \textbf{V}ariational \textbf{A}uto-\textbf{E}ncoder (CaD-VAE), a novel approach for learning causal disentangled representations from interaction data in recommender systems. 
The CaD-VAE method considers the causal relationships between semantically related factors in real-world recommendation scenarios, rather than enforcing independence as in existing disentanglement methods. The approach utilizes structural causal models to generate causal representations that describe the causal relationship between latent factors.
The results demonstrate that CaD-VAE outperforms existing methods, offering a promising solution for disentangling complex user behavior data in recommendation systems.
\end{abstract}




\maketitle

\section{Introduction}
Recommender systems play a crucial role in providing personalized recommendations to users based on their behavior. Learning the representation of user preference from behavior data is a critical task in designing a recommender model.
Recently, deep neural networks
have demonstrated their effectiveness in representation learning in recommendation models~\cite{zhang2019deep}.
However, the existing methods for learning representation from users’ behavior in recommender systems face several challenges. One of the key issues is the inability to disentangle the latent factors that influence users’ behavior. This often leads to a highly entangled representation, 
which would disregard the complex interactions between latent factors driving users’ decision-making.

Disentangled Representation Learning (DRL) has gained increasing attention 
as a proposed solution to tackle existing challenges~\cite{bengio2013representation}. 
Most research on DRL has been concentrated on computer vision~\cite{higgins2017beta, kim2018disentangling, dupont2018learning, liu2021smoothing, yang2021causalvae}, with few studies examining its applications in recommender systems~\cite{wang2022disentangled}. Recent works, such as CasualVAE~\cite{yang2021causalvae} and DEAR~\cite{shen2022weakly} utilize weak supervision to incorporate causal structure into disentanglement, allowing for the generation of images with causal semantics.

Implementing DRL in recommendation systems can enable a more fine-grained analysis of user behavior, leading to more accurate recommendations.
One popular DRL method is Variational Autoencoders (VAE)
~\cite{higgins2017beta, kumar2017variational, kim2018disentangling, quessard2020learning}, 
which learns latent representations capturing the data's underlying structure.
~\citet{ma2019learning} propose MacridVAE to learn the user’s macro and micro preference on items for
collaborative filtering. 
~\citet{wang2022disentangledb} extend the MacridVAE by employing visual images and textual descriptions to extract user interests. However, these works assume that countable, independent factors generate real-world observations, which may not hold in all cases. We argue that latent factors with the semantics of interest, known as concepts~\cite{ma2019learning, yang2021causalvae}, have causal relationships in the recommender system.
For example, in the movie domain, different directors specialize in different film genres, and different film genres may have a preference for certain actors.
As a result, learning causally disentangled representations reflecting the causal relationships between high-level concepts related to user preference would be a better solution.


In this work, we propose a novel approach for disentanglement representation learning in recommender systems by adopting a structural causal model, named Causal Disentangled Variational Auto-Encoder (CaD-VAE).
Our approach integrates the causal structure among high-level concepts that are associated with user preferences (such as film directors and film genres) into DRL. Regularization is applied to ensure each dimension within a high-level concept captures an independent, fine-grained level factor (such as action movies and funny movies within the film genre).
Specifically, the input data is first processed by an encoder network, which maps it to a lower-dimensional latent space, resulting in independent exogenous factors. The obtained exogenous factors are then passed through a causal layer to be transformed into causal representations, where additional information is used to recover the causal structure between latent factors.
Our main contributions are summarized as follows, 
\begin{itemize}
    \item  We introduce the problem of causal disentangled representation learning for sparse relational user behavior data in recommender systems.
    \item We propose a new framework named CaD-VAE, which is able to describe the SCMs for latent factors in representation learning for user behavior.
    \item We conduct extensive experiments on various real-world datasets, demonstrating the superiority of CaD-VAE over existing state-of-the-art models.
\end{itemize}

\section{Methodology}
\begin{figure}[h]
    \centering
    \includegraphics[width=\linewidth]{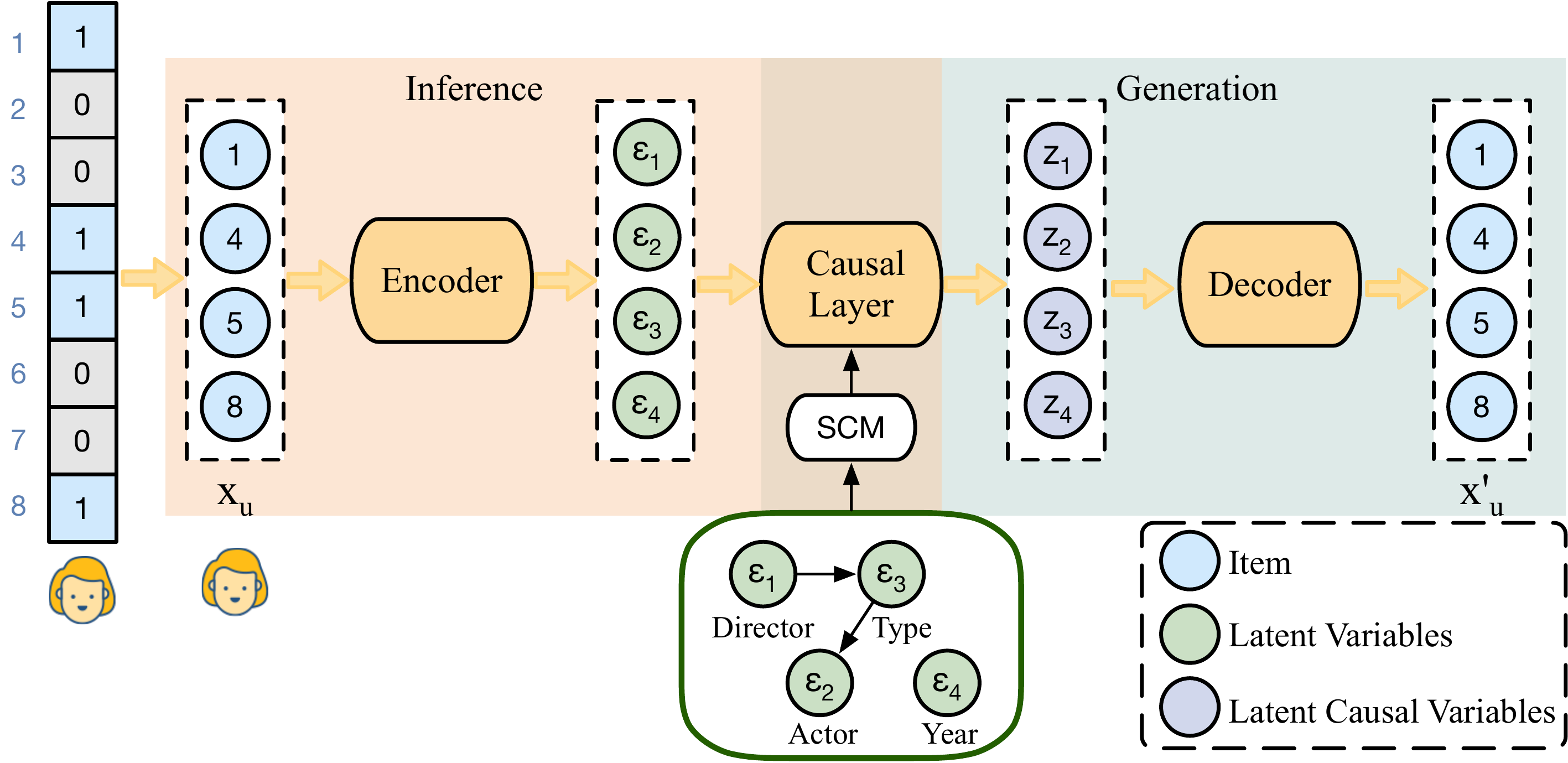}
    \caption{Model structure of CaD-VAE. 
    The encoder takes the observation $x_u$ as input to generate an independent exogenous variable $\epsilon$, which is then transformed into causal representations $z$ by the Causal Layer. The decoder uses $z$ as input to reconstruct the original observation $x_u$.
    }
    \label{fig:overview}
\end{figure}

In this section, we propose the \textbf{Ca}usal \textbf{D}isentangled \textbf{V}ariational \textbf{A}uto-\textbf{E}ncoder (CaD-VAE) method for causal disentanglement learning for the recommender systems. 
The overview of our proposed CaD-VAE model structure is in~\Cref{fig:overview}.

\subsection{Problem Formulation}
Let $u \in \{1,....,U\}$ and $i \in \{1,....,I\}$ index users and items, respectively. For the recommender system, the dataset $\mathcal{D}$ of user behavior consists of $U$ users and $I$ items interactions.
For a user $u$, the historical interactions 
$D_u = \{x_{u,i}: x_{u,i} \in \{0, 1\}\}$ 
is a multi-hot vector, where $x_{u,i} = 0$ represents that there is no recorded interaction between  user $u$ and item $i$, and $x_{u,i} = 1$ represents an interaction between the user $u$ and item $i$, such as click.
For notation brevity, we use $\mathbf{x}_u$ denotes all the interactions of the user $u$, that is $\mathbf{x}_u = \{x_{u,i}: x_{u,i} = 1 \}$.
Users may have very diverse interests, and interact with items that belong to many high-level concepts, such as preferred film directors, actors, genres, and year of production.
We aim to learn disentangled representations from user behavior that reflect the user preference related to different high-level concepts and reflect the causal relationship between these concepts.

\subsection{Construct Causal Structure via SCMs}
Consider $k$ high-level concepts in observations to formalize causal representation. These concepts are thought to have causal relationships with one another in a manner that can be described by a Directed Acyclic Graph (DAG), which can be represented as an adjacency matrix, denoted by $A$.
To construct this causal structure, we introduce a causal layer in our framework. This layer is specifically designed to implement the nonlinear Structural Causal Model (SCM) as proposed by~\cite{yu2019dag}:
\begin{equation}
\label{SCM_2}
\mathbf{z} = g((\mathbf{I-A^\top})^{-1}\mathbf{\epsilon}) := F_\alpha(\epsilon),
\end{equation}
where $A$ is the weighted adjacency matrix among the k elements of z, $\epsilon$ is the exogenous variables that $\epsilon \sim \mathcal{N}(0, I)$, $g$ 
is nonlinear element-wise transformations. The set of parameters of $A$ and $g$ are denoted by $\alpha = (A, g)$.
Additionally, $A_{ij}$ is non-zero if and only if $[z]_i$ is a parent of $[z]_j$, and the corresponding binary adjacency matrix is denoted by $I_A = I(A\neq 0)$, where $I(\cdot)$ is an element-wise indicator function. 
When $g$ is invertible,~\Cref{SCM_2} can be rephrased as follows:
\begin{align}
\label{SCM_2}
g^{-1}_i(z_i) = \mathbf{A_i^\top} g^{-1}_i(\mathbf{z}) + \epsilon_i,
\end{align}
which implies that after undergoing a nonlinear transformation of $g$, the factors $\mathbf{z}$ satisfy a linear SCM.
To ensure disentanglement, 
we use the labels of the concepts to be additional information, denoted as $c$.
The additional information $c$ is used to learn the weights of the non-zero elements in the prior adjacency matrix, which represent the sign and scale of causal effects. 

\subsection{Causality-guided Generative Modeling}
Our model is built upon the generative framework of the Variational Autoencoder (VAE) and introduces a causal layer to describe the SCMs.
Let $E$ and $D$ denote the encoder and decoder, respectively. We use $\theta$ to denote the set $(E, D, \alpha, \lambda)$ that contains all the trainable parameters of our model. 
For a user $u$, our generative model parameterized by $\theta$ assumes that the observed data are generated from the following distribution:
\begin{equation}
p_\theta(\mathbf{x}_u) = \mathbb{E}_{p_\theta(\mathbf{c})}\Big[\iint p_\theta(\mathbf{x}_u|\mathbf{\epsilon}, \mathbf{z}_u, \mathbf{c})p_\theta(\mathbf{\epsilon}, \mathbf{z}_u |\mathbf{c}) d\mathbf{\epsilon} d\mathbf{z}_u\Big],
\end{equation}
where
\begin{equation}
p_\theta(\mathbf{x}_u|\mathbf{\epsilon, z, c}) = \prod_{{x_{u,i}}\in\mathbf{x}_u } p_\theta(x_{u,i}|\mathbf{\epsilon, z, c}).
\end{equation}


Assuming the encoding process $\mathbf{\epsilon} = E(\mathbf{x}_u, \mathbf{c}) + \zeta$,
where $\zeta$ is the vectors of independent noise with probability density $p_\zeta$.
The inference models that take into account the causal structure can be defined as:
\begin{equation}
q_\phi(\mathbf{\epsilon}, \mathbf{z}_u |\mathbf{x}_u, \mathbf{c}) \equiv q(\mathbf{z}_u|\mathbf{\epsilon}) q_\zeta(\mathbf{\epsilon} - E(\mathbf{x}_u, \mathbf{c})),
\end{equation}
where $q_\phi(\mathbf{\epsilon}, \mathbf{z}_u |\mathbf{x}_u, \mathbf{c})$ is the approximate posterior distribution, parameterized by $\phi$, that models the distribution of the latent representation given the user's behavior data $\mathbf{x}_u$ and additional information $\mathbf{c}$. Since $\epsilon$ and $z$ have a one-to-one correspondence, we can simplify the variational posterior distribution as follows:
\begin{equation}
\begin{aligned}
\label{eq:posterior}
q_\phi(\mathbf{\epsilon}, \mathbf{z}_u |\mathbf{x}_u, \mathbf{c}) 
& = q(\mathbf{\epsilon}|\mathbf{x}_u, \mathbf{c})\delta(\mathbf{z}_u = F_\alpha(\epsilon)),\\
& = q(\mathbf{z}|\mathbf{x}_u, \mathbf{c})\delta(\mathbf{\epsilon} = F^{-1}_\alpha(\mathbf{z}_u)),
\end{aligned}
\end{equation}
where $\delta(\cdot)$ is the Dirac delta function. And we define the joint prior distribution for latent variables $\epsilon$ and $z$ as:
\begin{equation}
\label{eq:prior}
p_\theta(\epsilon, \mathbf{z}_u |\mathbf{c}) = p_\epsilon(\epsilon)p_\theta(\mathbf{z}_u |\mathbf{c}), 
\end{equation}
where $p_\epsilon(\epsilon) = \mathcal{N}(0, I)$ and $p_\theta(\mathbf{z}_u |\mathbf{c})$ is a factorized Gaussian distribution such that:
\begin{equation}
p_\theta(\mathbf{z}_u |\mathbf{c}) = \prod_{i=1}^n p_\theta(z_u^{(i)} |c_i), 
\end{equation}
where $p_\theta(z_u^{(i)} |c_i) = \mathcal{N}(\lambda_1(c_i), \lambda_2^2(c_i))$. $\lambda_1$ and $ \lambda_2$ are arbitrary functions.

Given the representation of a user, denoted by $\mathbf{z_u}$, the decoder's goal is to predict the item, out of a total of $I$ items, that the user is most likely to click. 
Let $D$ denotes the decoder. We assume the decoding processes $\quad \mathbf{x} = D(\mathbf{z}) + \xi$,
where $\xi$ is the vectors of independent noise with probability density $q_\xi$.
Then we define the generative model parameterized by parameters $\theta$ as follows:
\begin{equation}
p_\theta(\mathbf{x}_u|\mathbf{\epsilon, \mathbf{z}_u, c}) = p_\theta(\mathbf{x}_u|\mathbf{z}_u) \equiv p_\xi(\mathbf{x}_u-D(\mathbf{z}_u)). 
\end{equation}

\subsection{Disentanglement Objective}
Our objective is to learn the parameters $\phi$ and $\theta$ that maximize the evidence lower bound (ELBO) of $\Sigma_u\ln{p_\theta(\mathbf{x}_u)}$.
The ELBO is defined as the expectation of the log-likelihood with respect to the approximate posterior $q_\phi(\mathbf{\epsilon}, \mathbf{z}_u |\mathbf{x}_u, \mathbf{c})$, where $\mathbf{z}_u$ and $\epsilon$ are the latent variables:
\begin{equation}
\begin{aligned}
\label{eq:ELBO_1}
     ln{p_\theta(\mathbf{x}_u)}
     &\geq \mathbb{E}_{p_\theta(\mathbf{c})}[ \mathbb{E}_{q_\phi(\mathbf{\epsilon}, \mathbf{z}_u |\mathbf{x}_u, \mathbf{c})}[ln p_\theta (\mathbf{x}_u|\mathbf{\epsilon}, \mathbf{z}_u, \mathbf{c})] \\
     &- D_\mathrm{{KL}}(q_\phi(\mathbf{\epsilon}, \mathbf{z}_u |\mathbf{x}_u, \mathbf{c}) \| p_\theta(\mathbf{\epsilon}, \mathbf{z}_u | \mathbf{c}))],
\end{aligned}
\end{equation}
Based on the definitions of the approximate posterior in~\Cref{eq:posterior} and the prior distribution in~\Cref{eq:prior}, ELBO defined in~\Cref{eq:ELBO_1} can be expressed in a neat form as follows:
\begin{equation}
\begin{aligned}
\label{eq:ELBO_1}
     \text{ELBO} 
     &= \mathbb{E}_{p_\theta(\mathbf{c})}[ \mathbb{E}_{q_\phi(\mathbf{\mathbf{z}_u |\mathbf{x}_u, \mathbf{c})}}[ln p_\theta (\mathbf{x}_u|\mathbf{\epsilon}, \mathbf{z}_u, \mathbf{c})]\\
     &- D_\mathrm{{KL}}(q_\phi(\mathbf{\epsilon}, |\mathbf{x}_u, \mathbf{c}) \| p_\epsilon(\epsilon))\\
     &- D_\mathrm{{KL}}(q_\phi(\mathbf{z}_u |\mathbf{x}_u, \mathbf{c}) \| p_\theta(\mathbf{z}_u | \mathbf{c}))]
\end{aligned}
\end{equation}

Aside from disentangling high-level concepts, we are also interested in capturing the user's specific preference for fine-grained level factors within different concepts, such as action or funny films in the film genre.
Specifically, we aim to enforce statistical independence between the dimensions of the latent representation, so that each dimension describes a single factor, which can be formulated as forcing the following:
\begin{equation}
q_\phi(\mathbf{z}_{u}^{(i)} |\mathbf{c}) = \prod_{j=1}^d q_\phi(z_{u,j}^{(i)} |c_i).
\end{equation}
Follow the idea in~\cite{ma2019learning} that $\beta$-VAE can be used to encourage independence between the dimensions.
By varying the value of $\beta$, the model can be encouraged to learn more disentangled representations, where each latent dimension captures a single, independent underlying factor~\cite{higgins2017beta}.
As a result, we amplify the regularization term by a factor of $\beta \gg 1$, resulting in the following ELBO:
\begin{equation}
\begin{aligned}
\label{eq:ELBO_1}
      &\mathbb{E}_{p_\theta(\mathbf{c})}[ \mathbb{E}_{q_\phi(\mathbf{\mathbf{z}_u |\mathbf{x}_u, \mathbf{c})}}[ln p_\theta (\mathbf{x}_u|\mathbf{\epsilon}, \mathbf{z}_u, \mathbf{c})]\\
     &- D_\mathrm{{KL}}(q_\phi(\mathbf{\epsilon}, |\mathbf{x}_u, \mathbf{c}) \| p_\epsilon(\epsilon))\\
     &- \beta D_\mathrm{{KL}}(q_\phi(\mathbf{z}_u |\mathbf{x}_u, \mathbf{c}) \| p_\theta(\mathbf{z}_u | \mathbf{c}))]
\end{aligned}
\end{equation}

To ensure the learning of causal structure and causal representations, we include a form of supervision during the training process of the model. The first component of this supervision involves utilizing the extra information $c$ to establish a constraint on the weighted adjacency matrix $A$. This constraint ensures that the matrix accurately reflects the causal relationships between the labels:
\begin{equation}
L_{sup}^a = \mathbb{E}_{q_{\mathcal{X}}} \|c - \sigma(A^\top c) \|_2^2 \leq \kappa_1, 
\end{equation}
where $\sigma(\cdot)$ is the sigmoid function and $\kappa_1$ is a small positive constant value.
The second component of supervision constructs a constraint on learning the latent causal representation $z$:
\begin{equation}
L_{sup}^z = \mathbb{E}_{z \sim q_\phi} \Sigma_{i=1}^n \| g^{-1}_i(z_i) - \mathbf{A_i^\top} g^{-1}_i(\mathbf{z})\|_2^2 \leq \kappa_2, 
\end{equation}
where $\kappa_2$ is a small positive constant value. Therefore, we have the following training objective:
\begin{equation}
\mathcal{L} = -ELBO + \gamma_1L_{sup}^a+\gamma_2 L_{sup}^z,
\end{equation}
where $\gamma_1$ and $\gamma_2$ are regularization hyperparemeters.

\begin{table*}[ht]
    \centering
    \caption{Result comparison of the proposed method with several exciting works. The best results are bold, and the second best are marked as *.  All methods are constrained to have around $2M d$ parameters, where $d=100$.}
    \begin{tabular}{cccccc}
        Dataset & Method & NDCG@50 & NDCG@100 & Recall@20 &  Recall@50 \\ \hline
        ML-100k & MultDAE & 0.13226 (±0.02836) &  0.24487 (±0.02738)&  0.23794 (±0.03605)&  0.32279 (±0.04070)\\
        & $\beta$-MultVAE & 0.13422 (±0.02341)& 0.27484 (±0.02883) & 0.24838 (±0.03294) & 0.35270 (±0.03927) \\
        & MacridVAE & 0.14272 (±0.02877) & 0.28895 (±0.02739) & 0.30951 (±0.03808)* &  \textbf{0.41309 (±0.04503)}\\
        & DGCF & 0.15215 (±0.03612) & 0.28229 (±0.02271) & 0.28912 (±0.03012) & 0.34233 (±0.02937) \\
         & SEM-MacridVAE & 0.17322 (±0.02812)* & 0.29372 (±0.02371)* & 0.27492 (±0.02152) & 0.37026 (±0.02914) \\
         & Ours & \textbf{0.19272 (±0.02515)} & \textbf{0.31826 (±0.02018} & \textbf{0.31272 (±0.02612)} & 0.38162 (±0.03812)* \\\hline
         ML-1M & MultDAE & 0.29172 (±0.00729) & 0.40453 (±0.00799)&  0.34382 (±0.00961) & 0.46781 (±0.01032) \\
         & $\beta$-MultVAE & 0.30128 (±0.00617) & 0.40555 (±0.00809) & 0.33960 (±0.00919) & 0.45825 (±0.01039)\\
         & MacridVAE & 0.31622 (±0.00499) & 0.42740 (±0.00789) & 0.36046 (±0.00947) & 0.49039 (±0.01029)\\
         & DGCF & 0.32111 (±0.01028) & 0.43222 (±0.00617) & 0.37152 (±0.00891) & 0.49285 (±0.09918)  \\
         & SEM-MacridVAE & 0.32817 (±0.00916)* & 0.44812 (±0.00689)* & 0.38172 (±0.00798)* & 0.49871 (±0.01029)* \\
         & Ours & \textbf{0.34716 (±0.00718)} & \textbf{0.45971 (±0.00610)} & \textbf{0.39182 (±0.00571)} & \textbf{0.50127 (±0.00917)}\\\hline 
         ML-20M & MultDAE &0.32822 (±0.00187) & 0.41900 (±0.00209) &  0.39169 (±0.00271) & 0.53054 (±0.00285) \\
         & $\beta$-MultVAE & 0.33812 (±0.00207) & 0.41113 (±0.00212) & 0.38263 (±0.00273)&  0.51975 (±0.00289)\\
         & MacridVAE & 0.34918 (±0.00271) & 0.42496 (±0.00212) & 0.39649 (±0.00271) & 0.52901 (±0.00284) \\
         & DGCF & 0.36152 (±0.00281) & 0.43172 (±0.00199) & 0.40127 (±0.00284) & 0.52127 (±0.00229) \\
         & SEM-MacridVAE & 0.37172 (±0.00187)* & 0.44312 (±0.00177)* & 0.41272 (±0.00300)* & 0.53212 (±0.00198)*\\
         & Ours & \textbf{0.38991 (±0.00201)} & \textbf{0.45126 (±0.00241)}  & \textbf{0.42822 (±0.00298)} & \textbf{0.54316 (±0.00189)}\\\hline 
         Netflix & MultDAE & 0.24272 (±0.00089)& 0.37450 (±0.00095) & 0.33982 (±0.00123) & 0.43247 (±0.00126) \\
         & $\beta$-MultVAE & 0.24986 (±0.00080) & 0.36291 (±0.00094) & 0.32792 (±0.00122) & 0.41960 (±0.00125) \\
         & MacridVAE &0.25717 (±0.00098) & 0.37987 (±0.00096) & 0.34587 (±0.00124) & 0.43478 (±0.00118) \\
         & DGCF & 0.27128 (±0.00089)* & 0.39122 (±0.00078)* & 0.36271 (±0.00199)* & 0.45019 (±0.00102)* \\
         & SEM-MacridVAE & 0.26981 (±0.00100) & 0.38012 (±0.00099) &0.35712 (±0.00162) & 0.44172 (±0.00102) \\
         & Ours & \textbf{0.29172 (±0.00080)} & \textbf{0.40021 (±0.00088)} & \textbf{0.38212 (±0.00062)} & \textbf{0.45918 (±0.00081)}\\\hline 
    \end{tabular}
    \label{tab:result}
\end{table*}
\section{Experiment}

\subsection{Experiment Setup}
Our experiments were conducted on four datasets, which included a combination of real-world datasets. The largescale Netflix Prize dataset and three MovieLens datasets of different scales (i.e., ML-100k, ML-1M, and ML-20M) were used following the same methodology as MacridVAE. To binarize these four datasets, we only kept ratings of four or higher and users who had watched at least five movies. We choose four causally related concepts: (DIRECTOR $\rightarrow$ FILM GENRE), (FILM GENRE $\rightarrow$ ACTOR), (PRODUCTION YEAR).
We compare the proposed approach with four existing baselines:
\begin{itemize}
    \item MacridVAE~\cite{ma2019learning} is a disentangled representation learning method for the recommendation.
    \item $\beta$-MultVAE~\cite{liang2018variational} and MultiDAE~\cite{liang2018variational} are VAE based representation learning method for reommendation.
    \item DGCF~\cite{wang2020disentangled} is a disentangled graph-based method for collaborative filtering.
    \item SEM-MacridVAE~\cite{wang2022disentangledb} is the extension of MacridVAE by introducing semantic information. 
\end{itemize}
The evaluation metric used is nDCG and recall, which is the same as~\cite{wang2022disentangledb}. 
\subsection{Resutls}
\textbf{Overall Comparison.} 
The overall comparison can be found on~\Cref{tab:result}. We can see that the proposed method generally outperformed all of the existing works. It demonstrates that the proposed causal disentanglement representation works better than traditional disentanglement representation. 

\begin{figure}[h]
     \centering
     \begin{subfigure}[b]{0.48\linewidth}
         \centering
         \includegraphics[width=\linewidth]{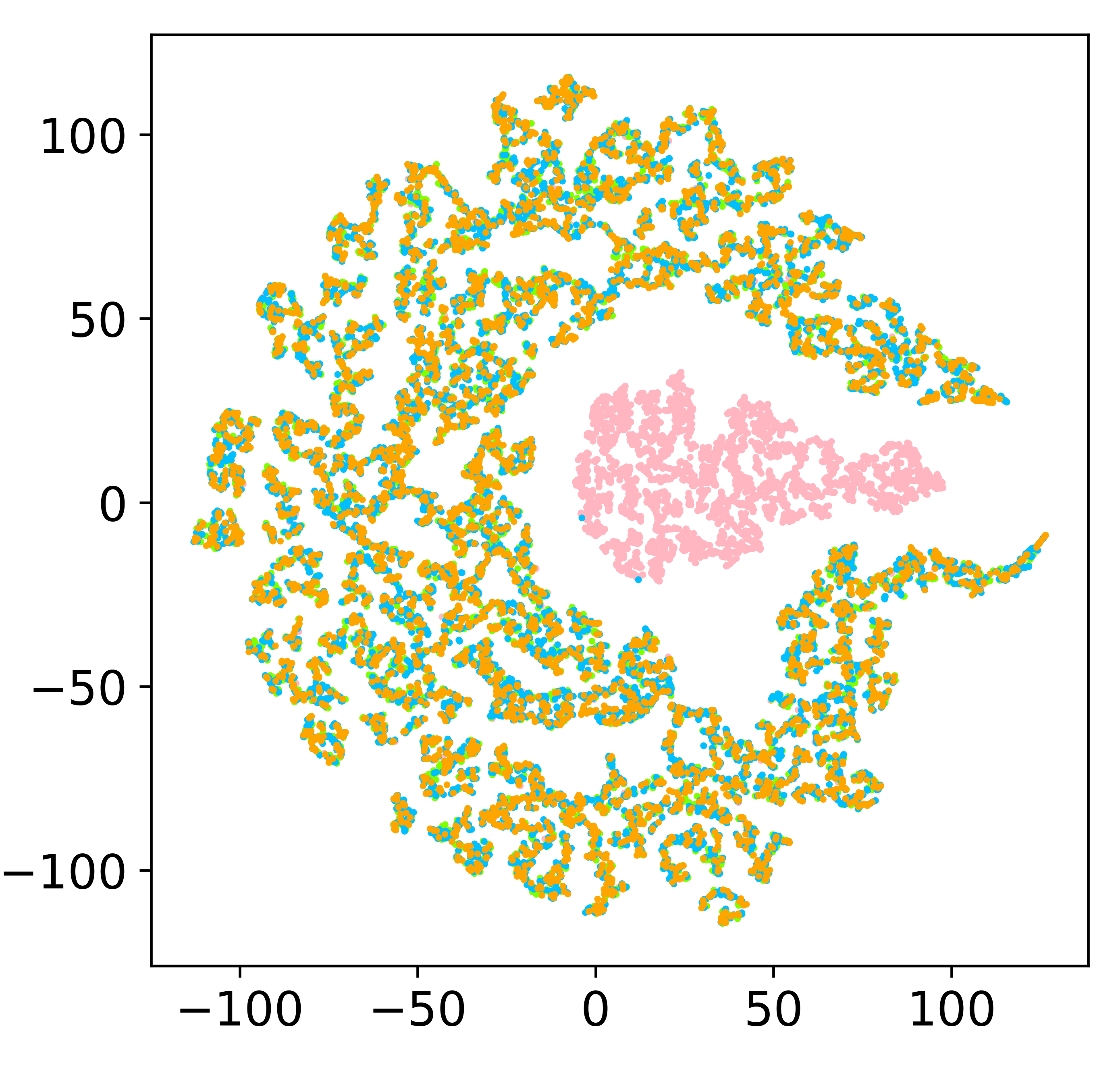}
         \caption{}
         \label{fig:visa}
     \end{subfigure}
     \begin{subfigure}[b]{0.498\linewidth}
         \centering
         \includegraphics[width=\linewidth]{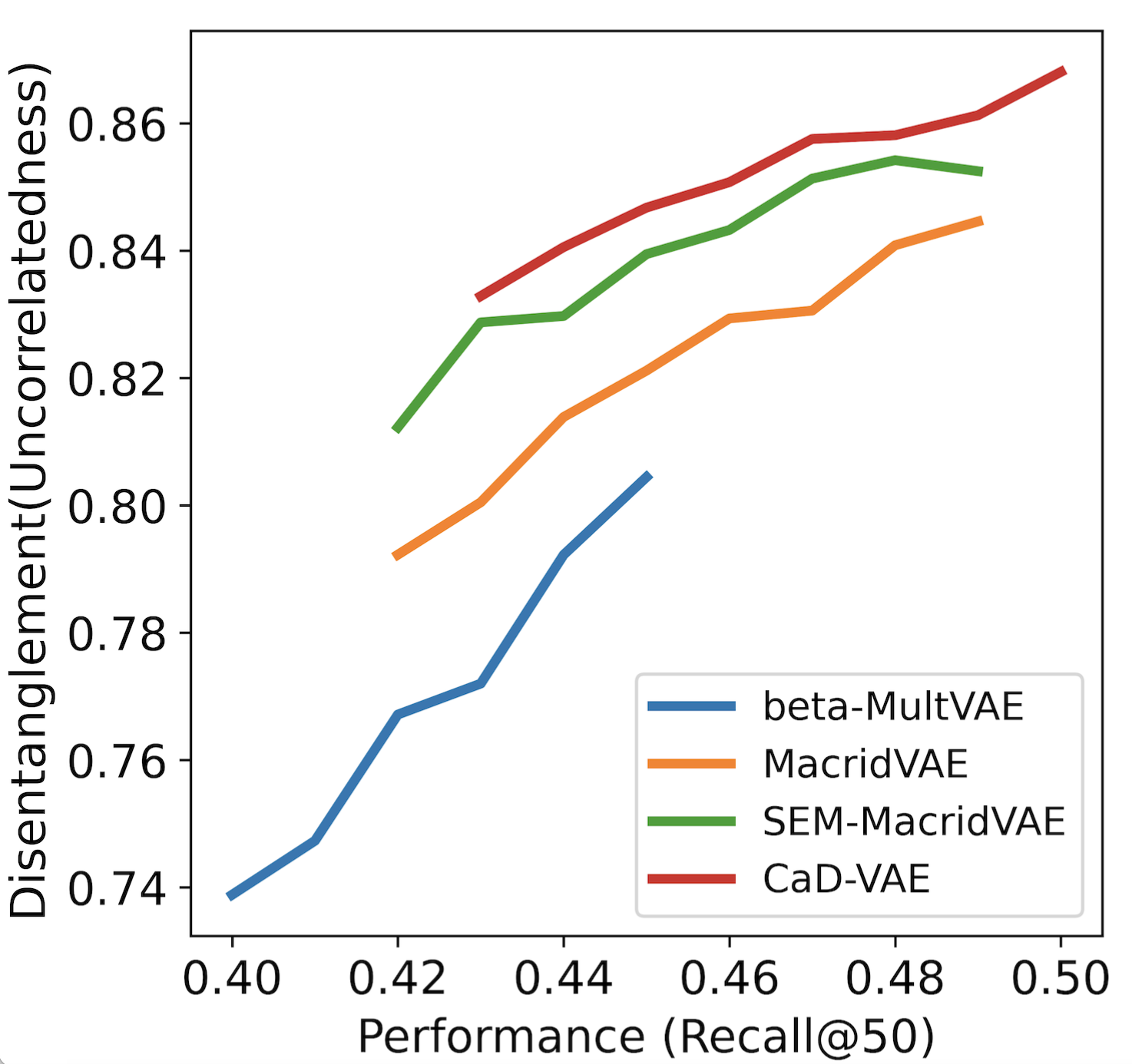}
         \caption{}
         \label{fig:dis}
     \end{subfigure}
        \caption{Disentanglement experiments. (a) is the visualization learned causal disentanglement representation; (b) reflects the impact of the fine-grained level disentanglement and the recommendation performance}
        \label{fig:disen}
\end{figure}

\noindent\textbf{Causal Disentanglement.}  We also provide a t-SNE~\cite{van2008visualizing} visualization of the learned causal disentanglement representation for high-level concepts on ML-1M. On the representation visualization~\Cref{fig:visa}, pink represents the year of the production, green represents the directors, blue represents the actors and yellow represents the genres. We can clearly find that the year of production is disentangled from actors, genres and directors as they are not causally related. 

\noindent\textbf{Fine-grained Level Disentanglement.} In \Cref{fig:dis}, we examine the relationship between the level of independence at the fine-grained level and the performance of recommendation by varying the hyper-parameter $\beta$. To quantify the level of independence, we use a set of $d$-dimensional representations and calculate the following metric
$1-\frac{2}{d(d-1)}\sum_{1\leq i<j \leq d}|\text{corr}_{i,j}|$~\cite{ma2019learning}, 
where $\text{corr}_{i,j}$ is the correlation between dimension $i$ and $j$. We observe a positive correlation between recommendation performance and the level of independence, where higher independence leads to better performance. Our method outperforms existing disentanglement representation learning in the level of independence.

\section{Conclusion}
This work demonstrates the effectiveness of the CaD-VAE model in learning causal disentangled representations from user behavior. 
Our approach incorporates a causal layer implementing SCMs, allowing for the successful disentanglement of causally related concepts. Experimental results on four real-world datasets demonstrate that the proposed CaD-VAE model outperforms existing state-of-the-art methods for learning disentangled representations. 
In terms of future research, there is potential to investigate novel applications that can take advantage of the explainability and controllability offered by disentangled representations.

\bibliographystyle{ACM-Reference-Format}
\bibliography{sample-base}
\end{document}